\documentclass[twocolumn,showpacs,preprintnumbers,amsmath,amssymb,superscriptaddress,10pt,aps,prb]{revtex4-2}
\usepackage{epsfig,amsopn}
\usepackage{graphicx}
\usepackage{epstopdf}
\usepackage{mathrsfs}  
\usepackage{amsmath,amssymb}
\usepackage{amsthm}
\usepackage{enumerate}
\usepackage{xcolor}
\usepackage[normalem]{ulem}
\usepackage[colorlinks=true,linktoc=page,linkcolor=magenta,citecolor=magenta]{hyperref}

\newcommand\bs{\boldsymbol}

\newcommand\bea{\begin{eqnarray}}
	\newcommand\eea{\end{eqnarray}}
\newcommand\beq{\begin{equation}}  
	\newcommand\eeq{\end{equation}}

\begin{document}
	\title{\bf{Operator correlations in a quenched non-Hermitian Luttinger liquid }}
	\author{Ashutosh Dubey}
        \email{adubey@iitk.ac.in}
	\author{Sourav Biswas}
        \email{souravbw@iitk.ac.in}
	\author{Arijit Kundu}
        \email{arijit.hri@gmail.com}
	\affiliation{Department of Physics, Indian Institute of Technology - Kanpur, Kanpur 208 016, India.}

\begin{abstract}
		
We study operator correlations of a spinful Luttinger liquid after introducing a non-Hermitian interaction quench, yielding supersonic modes and dominant superconducting correlations as signatures of the non-unitary dynamics as well as spin-charge separation. A comparative analysis with the Hermitian counterpart, i.e, when the quench is Hermitian, shows a significant difference in the behavior of the model. We derive exact expressions for different operator correlations and show that the superconducting correlations decay slower than the charge and spin-density wave correlations, especially, within the short-time limit, and at the long-time limit  all the operator correlations merge differed only by phase factors in the case of non-hermitian interaction quench whereas they do not merge in the case of Hermitian interaction quench. In both cases known Luttinger liquid universality is retained at the long time limit. We also analyze how the dynamics of operator correlations vary in the presence of anisotropy in the quenching parameters. 
		
\end{abstract}

\maketitle

\section{Introduction}


The low-energy physics of an interacting one-dimensional metallic system is generally described by the Luttinger liquid (LL) theory~\cite{GIAMARCHI, FRADKIN}, which is among the better-studied many-body quantum systems. At equilibrium, the nature of the system can be understood from various order parameter correlations, which signify tendencies to show quasi-long-range orders. One uses the bosonization technique in order to study the physics of LLs. The strength and usefulness of such a technique lie in its applicability to a wide range of circumstances \cite{GIAMARCHI,Citro2011}. This Luttinger liquid formalism has also been successfully implemented to study many out-of-equilibrium phenomena of one-dimensional interacting systems \cite{Cazalilla2006_1, Cazalilla2006_2, Dora2011, Dora2013_1, Dora2013_2, Meden2013, Dutta2014, Orignac2016, Sassetti2016, Chung2016}. 

Non-hermiticity in quantum mechanical systems has paved the way for several new phenomena \cite{Bender1998,Bender2006, Ashida2020,Sato2019}. Over the last few years, a diverse set of experiments have reported signatures of non-Hermitian physics \cite{Ghatak2020,Liu2022,Ding2021,Xiao2017,Weimann2017}. These developments have resulted in a growing interest in the theoretical study of non-Hermitian systems. Particular attention has been paid to the study of $\mathcal{PT}$-symmetric Hamiltonians as the eigenenergies of such Hamiltonians can become real \cite{Bender1998,Bender2006}. This peculiarity has been exploited by a significant number of works to study the many-body physics of $\mathcal{PT}$-symmetric systems \cite{Ganainy2018, Dora2020, Pereira2018, Das2021, Ashida2017}. Despite recent attempts~\cite{Ganainy2018, Dora2020, Pereira2018, Das2021, Ashida2017,Nakagawa2018,Lin2022,Hyart2022,Kawabata2022,Kazuki2022,Kazuki2019}, understanding of the consequences of non-Hermiticity in quantum many-body systems remains broadly unexplored.

The possibility of unveiling new physics by studying non-Hermitian systems using LL formalism has attracted the attention of researchers in recent times~\cite{Ashida2017,Kazuki2022,Dora2020}. The LL formalism provides analytical ease to study various many-body phenomena in non-Hermitian systems and becomes a natural tool to investigate the same, particularly in one dimension. Yamamoto et al. \cite{Kazuki2022} have studied dissipative LL originating from the non-Hermitian XXZ model. The authors have computed the correlation functions in detail where the right-state correlation functions take the same form as the order parameter correlations defined for hermitian spineless LL \cite{FRADKIN}. In another work,  Ashida et al. \cite{Ashida2017} have shown emergent superfluidity as a result of relevant non-Hermitian perturbation over the gapless model of Hermitian Luttinger liquid. Recently, LL formalism has been used to study non-Hermitian dynamics of many-body systems. In particular, Dora et al. \cite{Dora2020} investigated the single-particle Green's function and density-density correlator for non-Hermitian quantum quench in the LL where supersonic modes appear as a consequence of non-Hermiticity. Such modes are known to break the Lieb-Robinson bound \cite{LBB}. However, detailed studies and comparisons among different correlators are still absent in the literature, pertaining to non-Hermitian quantum quench dynamics of Luttinger liquids. Such analysis would require a more detailed study of operator correlations (OCs) as a function of time. We note that the application of OCs in the study of out-of-equilibrium physics is not new in the context of Hermitian LL \cite{Cazalilla2006_1,Cazalilla2006_2, Manmana2020, Moessner2017}. It is also very important to note that supersonic modes are not specific to non-Hermitian systems \cite{Kastner2013} and it is of general interest to enquire what lies beyond these modes.

In this work, we explore the properties of various correlators of a spinful LL model, under the influence of a sudden non-Hermitian interaction quench. We explicitly calculate different operator correlations for the quenched state and compare the result with the Hermitian counterpart. Interestingly, our results show that in the short time limit, superconducting correlations dominate over charge-density wave or spin-density wave correlations. The dominance of superconducting operator correlations is also observed for the spinless case. We further show how the dominant behavior varies with anisotropy in quenching parameters.

The organization of the paper is as follows: In Sec. \ref{Sec:Model} we discuss the model and the equation of motion for the bosonic operators. In Sec. \ref{Sec:BSM} we derive general expressions for various operator correlations required to study the system after quenching. We derive all quantities exactly to compare between Hermitian and non-hermitian quench in LLs and elaborate on the long-time limit of the operator correlations. In Sec. \ref{Sec:RD} we discuss the short-time limit of the operator correlations, the effect of anisotropy, and other crucial observations. In Sec. \ref{Sec:Sum} we summarize our work.
\section{Model} \label{Sec:Model}
A simple model describing a spinful LL can be written as \cite{Voit1995,Schon2003,Meden2016}:
\begin{equation}
\begin{aligned} \label{model}
	H= \sum_{p\ne 0, \nu} v |p| b_{p, \nu}^{\dagger}b_{p, \nu}+ \frac{i g_{2\nu} |p| \bs{\Theta}(t)}{2}\left(b_{p, \nu}^{\dagger}b_{-p, \nu}^{\dagger}+ b_{p, \nu}b_{-p, \nu}\right) ,
\end{aligned}	
\end{equation}
where \(\nu=c\) and \(\nu=s\) denote the charge and spin degrees of freedom, respectively. $\bs{\Theta}(t)$ is the unit-step function and \( i = \sqrt{-1} \). \(b_{p,\nu}\) is the annihilation operator for bosonic excitation of flavor $\nu$ and momentum $p$. $v$ is the bare sound velocity and $g_{2\nu}$ is the coupling constant, which is better understood  when viewed in the context of the underlying fermionic theory. In the fermionic language, $g_{2 \nu}$ denotes density-density interaction between the left and right movers. One can define $g_{2\nu} = g_{2 \mid \mid} \pm  g_{2 \perp}$, where the (+) sign holds for the charge sector ($\nu=c$) and the (-) sign holds for the spin sector ($\nu =s$) \cite{Voit1995, Schon2003}. Here $g_{2 \mid \mid}$ is the interaction between the left and right movers' density operators carrying the same spin, whereas, \( g_{2 \perp} \) is the interaction between the left and right movers' density operators carrying opposite spins. We note that difference in \( g_{2 \perp} \) and $g_{2 \mid \mid}$ results in unequal \(g_{2c}\) and \( g_{2s} \). This is also equivalent to introducing anisotropy in the quenching (interaction) parameters. We discuss in Sec. \ref{Sec:RD} that such anisotropies can dictate dominating behavior of the system.

Before the quench (i.e, $t<0$), the above Hamiltonian describes a non-interacting bosonic system. The interaction quench at $t=0$ has imaginary  coupling strength, which makes the Hamiltonian non-Hermitian. In spite of non-Hermiticity, the spectrum remains real at \(t > 0\) and is given by \( \tilde{v}_{\nu}| p |\), where \(\tilde{v}_{\nu}= \sqrt{v^{2}+ g_{2\nu}^{2}}\) represents the re-normalized velocity. Also, note that the system remains stable for the limit \( \mid g_{2 \nu} \mid  < v \), and in the rest of the paper we maintain this limit unless mentioned otherwise.

Since we are interested in the time evolution of the system, we first focus on writing down the equations governing the time evolution of the boson operators. In the presence of non-Hermiticity we work in the pseudo-Heisenberg picture \cite{Dora2020}, where the equations of motion for the bosonic operators $\{b_{p, \nu}\}$ and $\{ b^{\dagger}_{p, \nu} \}$ can be written as,
\begin{equation}
\begin{aligned} \label{PHTE}
	i\partial_{t} b_{p, \nu}(t)= \left[b_{p, \nu}(t),H\right],\\
	i\partial_{t} b_{p, \nu}^{\dagger}(t)= \left[b_{p, \nu}^{\dagger}(t),H\right].
\end{aligned}	
\end{equation}
General solutions of Eq.~(\ref{PHTE}) takes the form
\begin{equation}
\begin{aligned} \label{ansatz}
	b_{p, \nu}(t)= u_{p, \nu}(t)b_{p, \nu}+ v_{p, \nu}(t)b_{-p, \nu}^{\dagger}, \\
	b_{-p, \nu}^{\dagger}(t)= u_{p, \nu}^{*}(t)b_{-p, \nu}^{\dagger}- v_{p, \nu}(t)b_{p, \nu},
\end{aligned}	
\end{equation}
where $u_{p, \nu}(t)$ and $v_{p, \nu}(t)$ are the Bogoliubov coefficients, satisfying the constraint $|u_{p, \nu}(t)|^2 + |v_{p, \nu}(t)|^2 = 1$. As a result of non-Hermiticity, the Hermitian conjugate of the operator \(b_{p, \nu}(t)\) is not related to \(b_{-p, \nu}^{\dagger}(t)\). These coefficients take the form, 
\begin{equation}
\begin{aligned} \label{coeff}
	u_{p, \nu}(t)&= \cos(\tilde{v}_{\nu}| p | t)- \frac{i v}{\tilde{v}_{\nu}}\sin(\tilde{v}_{\nu}| p | t),\\
	v_{p, \nu}(t)&= \frac{g_{2\nu}}{\tilde{v}_{\nu}}\sin(\tilde{v}_{\nu} 
	| p | t). 
\end{aligned}	
\end{equation}
The above solutions satisfy the boundary conditions \(u_{p, \nu}(0) = 1\) and \(v_{p, \nu}(0) = 0\). For the Hermitian case, Bogoliubov coefficients satisfy the constraint $|u_{p, \nu}(t)|^2 - |v_{p, \nu}(t)|^2 = 1$, and $b_{p, \nu}(t)= u_{p, \nu}(t)b_{p, \nu}+ v^{\ast}_{p, \nu}(t)b_{-p, \nu}^{\dagger}$, such that $b^{\dagger}_{p, \nu}(t)$ is obtained by taking the Hermitian conjugate \cite{Dora2011,Dutta2014,Sassetti2016}. 

If we use the same mapping \( g_{2 \nu} \to - i g_{2 \nu}\) to map non-Hermitian Hamiltonian to Hermitian Hamiltonian, then the Bogoliubov coefficients are mapped to the same coefficients defined for a Hermitian quench problem but with a negative interaction strength. We have shown the exact expressions of these coefficients in Appendix \ref{app-A5} while discussing the details of Hermitian quench. 
\section{Beyond supersonic modes}\label{Sec:BSM}
In this section, we investigate the OCs of a spinful LL using the developments made in the last section. We study the operator correlations, as a function of time, computed over the quenched state $| \psi (t) \rangle  = e^{-i H t} | \psi_0 \rangle $ obtained by evolving the ground state $| \psi_0 \rangle$ of the non-interacting Hamiltonian defined by Eq. (\ref{model}) at $t \to 0^{-}$. It is noted that the time-evolution takes place by the Hamiltonian defined in the same equation but for $t \to 0^{+}$.

The operators that are being used to define these correlators can be any of the following: CDW (charge density wave), $\rm SDW_{x,y,z}$ (spin density wave), SS (singlet superconductor), $\rm TS_{x,y,z}$ (triplet superconductor). We use the symbol $\mathbb{O}$ to denote these operators and detailed expressions of these operators are furnished in Table. \ref{Tab}. One should note that here we are using the same operator identities used to characterize phases of spinful LL in equilibrium \cite{GIAMARCHI, FRADKIN} but the expectation values are computed out-of-equilibrium. 
The general form of the OCs, in the case of non-Hermitian interaction quench dynamics, is given as:
\begin{equation}
\begin{aligned} \label{OCNH}
	\langle \mathbb{O}^{\dagger}(x) \mathbb{O}(0) \rangle_{NH} 
	= \frac{\langle \psi_0| e^{iH^{\dagger}t} e^{-iHt} \mathbb{O}^{\dagger}(x,t) \mathbb{O}(0,t) | \psi_0\rangle}{\mathcal{N}(t)},
\end{aligned}	
\end{equation}
where $\mathcal{N}(t) = \langle \psi_0| e^{iH^{\dagger}t} e^{-iHt} | \psi_0\rangle$. Calculating any correlation involves computing the product of operators of the form $e^{iH^{\dagger}t} e^{-iHt}$, which reflects the non-unitarity of time evolution. We can write this product in a simpler manner using the symmetry of the system. The Hamiltonian contains operators of the form $\mathcal{K}_{p, \nu}^{+} = b^{\dagger}_{p, \nu} b^{\dagger}_{-p, \nu}$, $\mathcal{K}_{p, \nu}^{-} = b_{p, \nu} b_{-p, \nu}$, $\mathcal{K}_{p, \nu}^{o} = (b^{\dagger}_{p, \nu} b_{p, \nu} + b_{-p, \nu} b^{\dagger}_{-p, \nu})/2$. Here, $\mathcal{K}_{p, \nu}^{\pm,o}$ are the generators of SU(1,1) Lie group. The faithful matrix representation of SU(1,1) Lie group  \cite{SU211} can be used to write the evolution operator
\begin{figure*}[t]
	\centering
	\includegraphics[width=.8\textwidth]{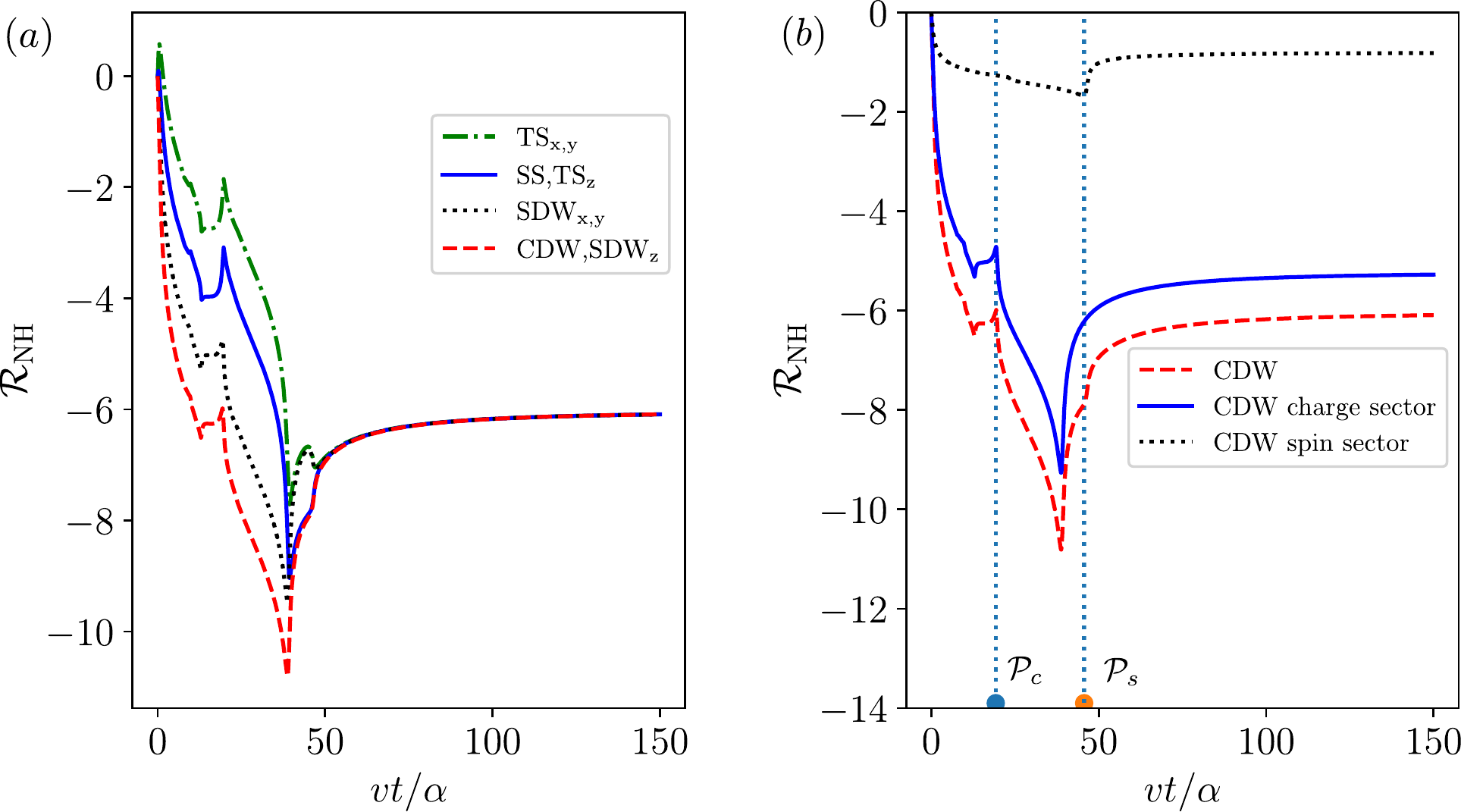}
	\caption{ Supersonic modes are a characteristic feature of non-Hermitian quench. It is expected to observe the appearance of such modes in the scaling of the OCs. However, comparison among different operator correlations shows the dominance of superconducting correlations over charge density wave or, spin density wave OCs, in short time limit. For the above two plots, used system parameters are: \(g_{2 \mid \mid}/v \) = 0.6, \(g_{2 \perp}/v \) = 0.2, \(x=100\alpha\). The sequence of the OCs is of particular importance, which would be clear in forthcoming sections of the paper. In ($a$) we plot exponent $\mathcal{R}_{NH}$ as a function of time. A larger value of exponent signifies slower decay and hence the dominating behavior of the system is governed by the operator having the largest value of $\mathcal{R}_{NH}$. The panel ($b$) shows how the information of spin-charge separation is being carried by the supersonic modes. The property of LL ensures that the OCs can be written as a product of charge and spin sectors. We show both the sectors separately and the combined sum using three different lines (for CDW-OC only). One can see different cusps coming from the charge sector and spin sector.  $\mathcal{P}_c$ denotes the point on the time axis where the cusp appears from the charge sector and, $\mathcal{P}_s$ denotes the cusp appearing from the spin sector. }\label{fig:NH}
\end{figure*}
\begin{equation}
		U_{\nu}(t)\label{Unut}
		=
		\prod_{p>0} e^{\mathcal{C}_{+ \nu}(p,t) \mathcal{K}_{p, \nu}^{+}} e^{\mathcal{C}_{0 \nu}(p,t) \mathcal{K}_{p, \nu}^{0}} e^{\mathcal{C}_{- \nu}(p,t) \mathcal{K}_{p, \nu}^{-}}, 
\end{equation}
where $\mathcal{C}_{n \nu}(p,t)$, with $n=\{0,\pm\}$, are some coefficients to be determined from the relation 
\begin{equation}\label{Ucond}
		U_{\nu}(t) = e^{i H^{\dagger}_{\nu} t} e^{- i H_{\nu} t},
\end{equation}
subject to the boundary conditions $\mathcal{C}_{n,\nu} (p,t=0) = 0 $, such that $U_{\nu}(0) = 1$. The total evolution operator is a product of $U_\nu(t)$ operators and it can be written in a compact form $e^{iH^{\dagger} t}e^{-i H t} = \prod_\nu U_\nu(t) $. 

In order to find the analytical expression of OCs, one requires to find the exact solutions of $\mathcal{C}_{n \nu}(p,t)$. It can be shown from Eq. (\ref{OCNH}) that the contribution due to $\mathcal{C}_{0 \nu}$ term cancels with $\mathcal{N}(t)$ and the contribution due to $\mathcal{C}_{+\nu}(p,t)$ term vanishes due to the fact that $b_{p, \nu}^{\dagger}$ annihilates $\langle \psi_o | $. As a result, one is left with the only requirement of computing $\mathcal{C}_{-\nu}(p,t)$. One can show from Eqs. (\ref{Unut}, \ref{Ucond}) that the differential equations governing the time evolution of $\mathcal{C}_{n \nu}(p,t)$ become 
\begin{equation}
\begin{aligned} 
	i \dot{\mathcal{C}}_{0 \nu}(p,t) &=  2 \beta_{p, \nu}(~ \mathcal{C}_{+\nu}(p,t) + \mathcal{C}_{-\nu}(p,t) ~), \\
	i \dot{\mathcal{C}}_{\pm \nu}(p,t) &=  \beta_{p, \nu}(1+e^{\mathcal{C}_{0\nu}(p,t)})  \mp \alpha_{p, \nu}\mathcal{C}_{\pm \nu}(p,t) \\&+ \beta_{p, \nu} \mathcal{C}^2_{\pm \nu}(p,t).  \\
\end{aligned}	
\end{equation} \\
We define $\alpha_{p, \nu} = 2 v |q| $ and $\beta_{p, \nu} = i g_{2 \nu} |q| $. Solving these equations one can obtain the solutions for $\mathcal{C}_{-\nu}(p,t)$ as,
\begin{equation}
	\begin{aligned} \label{Cm}
		\mathcal{C}_{-\nu}= \frac{2u_{p, \nu}(t)v_{p, \nu}(t)}{\mid u_{p, \nu}(t)\mid^{2}- \mid v_{p, \nu}(t)\mid^{2}}.
	\end{aligned}	
\end{equation}
We use these solutions to calculate the OCs for different operators. 

The $\mathbb{O}$ operators which are furnished in Table. \ref{Tab} require the knowledge of the dual boson fields $\phi_{\nu}(x)$ and $\theta_{\nu}(x)$. Below we mention the relation between $b_{p, \nu}$ operators and the field operators \cite{GIAMARCHI,Voit1995}
\begin{widetext}
\begin{equation}
\begin{aligned} \label{Dual_Fld}
	\phi_{\nu}(x)&= -i\frac{\pi}{L}\sum_{p>0}\left(\frac{L p}{2 \pi}\right)^{\frac{1}{2}}\frac{1}{p}e^{-\frac{\alpha p}{2}}\left(e^{-i p x}b_{p, \nu}^{\dagger}- e^{i p x}b_{-p, \nu}^{\dagger}+ e^{-i p x}b_{-p, \nu}- e^{i p x}b_{p, \nu}\right),
	\\
	\theta_{\nu}(x)&= i\frac{\pi}{L}\sum_{p>0}\left(\frac{L p }{2\pi}\right)^{\frac{1}{2}}\frac{1}{p}e^{-\frac{\alpha p}{2}}\left(e^{-i p x}b_{p, \nu}^{\dagger}+ e^{i p x}b_{-p, \nu}^{\dagger}- e^{-i p x}b_{-p, \nu}- e^{i p x}b_{p, \nu}\right).
\end{aligned} 
\end{equation}
Here $L$ is the length of the system, although we calculate the final expressions in the thermodynamic limit. $\alpha$ is an inherent length scale associated with the underlying lattice.
\end{widetext}
In the present work, the time dependence of the exponent $\mathcal{R}_{NH}$, in particular, is of our interest. We define $\mathcal{R}_{NH}$, which captures the scaling of the OCs, 
\begin{align}
	\mathcal{R}_{NH} = \log \Big(  \frac{\langle \mathbb{O}^{\dagger}(x) \mathbb{O}(0) \rangle_{NH}}{ \langle \mathbb{O}^{\dagger}(x) \mathbb{O}(0) \rangle_{0} } \Big).
\end{align}
We note that the exponent $\mathcal{R}_{NH}$ is a measure of the rate of decay of the OCs. We have found the exponent to be real for all the OCs. The term $\langle \mathbb{O}^{\dagger}(x) \mathbb{O}(0) \rangle_{0}$ denotes the operator correlation of the bare non-interacting system and it varies for different OCs by phase factors. In Appendix \ref{app-A1}, we elaborate on important expressions, used to compute all the operator correlations mentioned in Table. \ref{Tab}. The methods for calculating all the OCs follow from the derivations of $\langle \mathbb{O}^{\dagger}_{CDW}(x) \mathbb{O}_{CDW}(0) \rangle_{NH}$ and  $\langle \mathbb{O}^{\dagger}_{SS}(x) \mathbb{O}_{SS}(0) \rangle_{NH}$ respectively.  We refer to Appendix \ref{app-A2} for the details and report the general expression of OCs for non-Hermitian quantum quench in spinful LL below. 
\begin{widetext}	
	\begin{equation}
		\begin{aligned}
			\frac{\langle \mathbb{O}^{\dagger}(x) \mathbb{O}(0) \rangle_{NH}}{\langle \mathbb{O}^{\dagger}(x) \mathbb{O}(0) \rangle_{0}}
			=
			&\exp\left(\left(1-\frac{\tilde{v}_{c}}{v_{-c}}\right)D_{c}(x,0)- 2\sum_{n=1}^{\infty}\frac{\tilde{v}_{c}}{v_{-c}}\left(\frac{-g_{2c}^{2}}{v^{2}+ \tilde{v}_{c}v_{-c}}\right)^{n} D_{c}(x,nt) \right) \\    
			\times &
			\exp\left(\Delta_{c}\frac{g_{2c}}{\tilde{v}_{c}}\sum_{n=0}^{\infty}\sum_{m=1}^{n+1}(-1)^{m} \mathcal{M}_{n,m} \left(\frac{g_{2c}^{2}}{2\tilde{v}_{c}^{2}}\right)^{n}  F_{c}(x,mt)\right) \\
			\times &
			\exp\left(\left(1-\frac{\tilde{v}_{s}}{v_{-s}}\right)D_{s}(x,0)- 2\sum_{n=1}^{\infty}\frac{\tilde{v}_{s}}{v_{-s}}\left(\frac{-g_{2s}^{2}}{v^{2}+ \tilde{v}_{s}v_{-s}}\right)^{n}D_{s}(x,nt)  \right) \\   
			\times & \exp\left(\Delta_{s}\frac{g_{2s}}{\tilde{v}_{s}}\sum_{n=0}^{\infty}\sum_{m=1}^{n+1}(-1)^{m}\mathcal{M}_{n,m} \left(\frac{g_{2s}^{2}}{2\tilde{v}_{s}^{2}}\right)^{n}  F_{s}(x,mt)\right).
		\end{aligned}\label{eq:NH_gen}
	\end{equation}
Here,  $\mathcal{M}_{n,m} = {}^{2n}\mathbf{C}_{n-m-1}- {}^{2n}\mathbf{C}_{n-m+1}$, where $\mathbf{C}$ is used for combination. $\tilde{v}_{\nu}= \sqrt{v^{2}+ g_{2\nu}^{2}}$, $ v_{-\nu}= \sqrt{v^{2}- g_{2\nu}^{2}}$, $\omega_{p,\nu}= \tilde{v}_{\nu}\mid p\mid$,  $u_{p,\nu}(t)= \cos(\omega_{p,\nu}t)- \frac{iv}{\tilde{v}_{\nu}}\sin(\omega_{p,\nu}t)$, $v_{p,\nu}(t)= \frac{g_{2\nu}}{\tilde{v}_{\nu}}\sin(\omega_{p,\nu}t)$.  \(D_{\nu}(x,t)= \frac{1}{4}\left(\log\left(1+ \frac{x^{2}}{(\alpha- 2i\tilde{v}_{\nu}t)^{2}}\right)+ \log\left(1+ \frac{x^{2}}{(\alpha+ 2i\tilde{v}_{\nu}t)^{2}}\right)\right)\).  \(F_{\nu}(x,t)= \frac{1}{4i}\left[\log\left(1+ \frac{x^{2}}{(\alpha- 2i\tilde{v}_{\nu}t)^{2}}\right)- \log\left(1+ \frac{x^{2}}{(\alpha+ 2i\tilde{v}_{\nu}t)^{2}}\right)\right]\) . We recall $\nu=c,s$.
\end{widetext}
Different operator correlations can be obtained from Eq. (\ref{eq:NH_gen}) for different values of $\Delta_{c,s}$. We provide a dictionary in Table. \ref{Tab}. In Fig. \ref{fig:NH}  we have shown that in the short time limit for non-Hermitian quench, different cusp corresponds to the supersonic modes, and the position of one cusp is dominated by the contribution coming from either the charge sector or the spin sector. This is because of spin-charge separation. 
\begin{figure*}
	\centering
	\includegraphics[width=.8\textwidth]{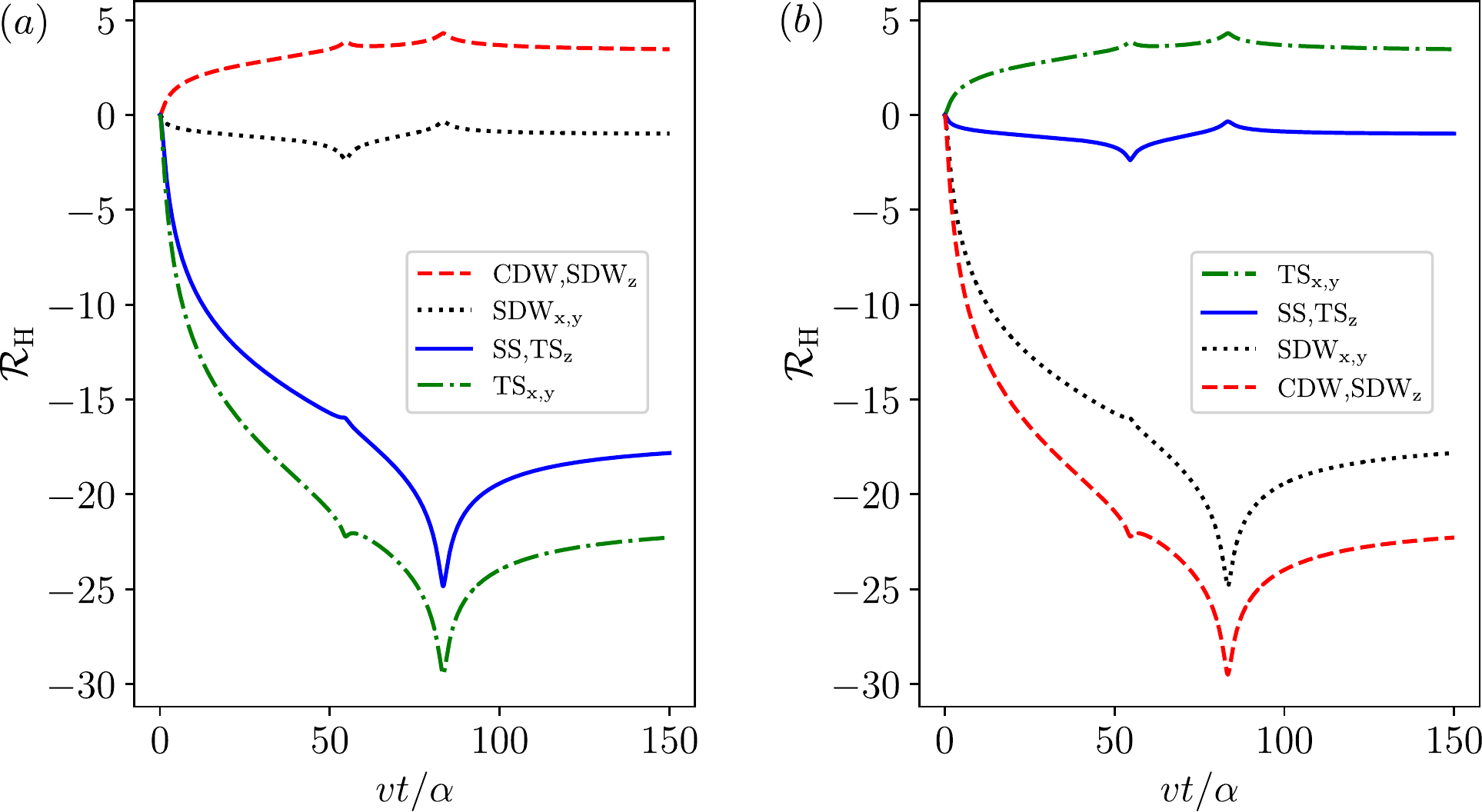}
	\caption{In ($a$) we show different OCs for various $\mathbb{O}$'s in case of Hermitian quench. One can observe that the CDW operator has the slowest decay, signifying the repulsive nature of the system. For this plot used system parameters are: \(g_{2 \mid \mid}/v \) = 0.6 , \(g_{2 \perp}/v \) = 0.2 , \(x=100\alpha\). ($b$) However, one can change the sign of the function \(g_{2 \mid \mid}\) to \(-g_{2 \mid \mid} \), and \(g_{2 \perp}\) to  \(-g_{2 \perp} \), to see the superconducting correlations becoming dominant. This is expected due to introducing negative interaction through quench. One must notice the alteration in the sequence of the OCs, between ($a$) and ($b$). }\label{fig:Her}
\end{figure*}
In the long time limit the time-dependent part $D_{\nu}(x,t)$ and $F_{\nu}(x,t)$ vanishes. The long time behavior of OCs is given by,
\begin{align}\label{eq:NH_gen_LT}
	\frac{\langle \mathbb{O}^{\dagger}(x) \mathbb{O}(0) \rangle_{NH}}{\langle \mathbb{O}^{\dagger}(x) \mathbb{O}(0) \rangle_{0}} \Bigg|_{t\to \infty}=~
	\nonumber
	&\exp\left(\left(1-\frac{\tilde{v}_{c}}{v_{-c}}\right)D_{c}(x,0)\right)\\
	\times & \exp\left(\left(1-\frac{\tilde{v}_{s}}{v_{-s}}\right)D_{s}(x,0)\right). 
\end{align}
At long time, one can not distinguish between different OCs from $\mathcal{R}_{NH}$. In the following segment of the paper, we contrast this behavior with the Hermitian case. 
\subsubsection*{Hermitian case}
One can study the Hermitian case in a manner similar to the non-Hermitian quench. We furnish the details in Appendix \ref{app-A5}. In this case, no supersonic modes appear and the Lieb-Robinson bounds are obeyed. One expects two cusps as shown in Fig. \ref{fig:Her}, one coming from the spin sector and another from the charge sector. In the long time limit the time-dependent function $D_{\nu}(x,t)$  vanishes. Whereas the time-independent part is different for CDW and SS, and they do not merge asymptotically.

For Hermitian case the normalization factor $\mathcal{N}(t)$ is unity and the correlation function reduces to $ \langle \psi_0| \mathbb{O}^{\dagger}(x,t) \mathbb{O}(0,t) | \psi_0\rangle $. In this case, the general expression for operator correlation is given by
\begin{align}\label{eq:H_gen}
	&\frac{\langle \mathbb{O}^{\dagger}(x) \mathbb{O}(0) \rangle_{H}}{\langle \mathbb{O}^{\dagger}(x) \mathbb{O}(0) \rangle_{0}} \nonumber \\=&  \exp \left(-\frac{g_{2c}(g_{2c}+ \Delta_{c}v)}{v_{-c}^{2}}(D_{c}(x,0)- D_{c}(x,t) ) \right) \nonumber \\
	\times &  \exp \left( - \frac{g_{2s}(g_{2s}+ \Delta_{s}v)}{v_{-s}^{2}} (D_{s}(x,0)-D_{s}(x,t)) \right),
\end{align}
where \(v_{-\nu}= \sqrt{v^{2}- g_{2\nu}^{2}}\).  We define the exponent as, $\mathcal{R}_{H} = \log \Big( \langle \mathbb{O}^{\dagger}(x) \mathbb{O}(0) \rangle_{H} / \langle \mathbb{O}^{\dagger}(x) \mathbb{O}(0) \rangle_{0} \Big) $. And the long time behavior of operator correlation is given by
\begin{align}\label{eq:H_gen_LT}
	\frac{\langle \mathbb{O}^{\dagger}(x) \mathbb{O}(0) \rangle_{H}}{\langle \mathbb{O}^{\dagger}(x) \mathbb{O}(0) \rangle_{0}} \Bigg|_{t\to \infty}
	\nonumber
	=&\exp \left(-\frac{g_{2c}(g_{2c}+ \Delta_{c}v)}{v_{-c}^{2}} D_{c}(x,0) \right) \\
	\times & \exp \left(-\frac{g_{2s}(g_{2s}+ \Delta_{s}v)}{v_{-s}^{2}} D_{s}(x,0) \right).
\end{align}
The expressions can be calculated for various values of $\Delta_{c,s}$, as done for the non-Hermitian case. In this case, as well, $\Delta_{c,s}$ follow from Table. \ref{Tab}. It is clear from the above expression that at the $t \to \infty$ limit, different correlations are distinguishable. We note that the computation of various OCs for hermitian quench is similar to that of the non-Hermitian counterpart. 
\begin{figure*}
	\centering
	\includegraphics[width=.8\textwidth]{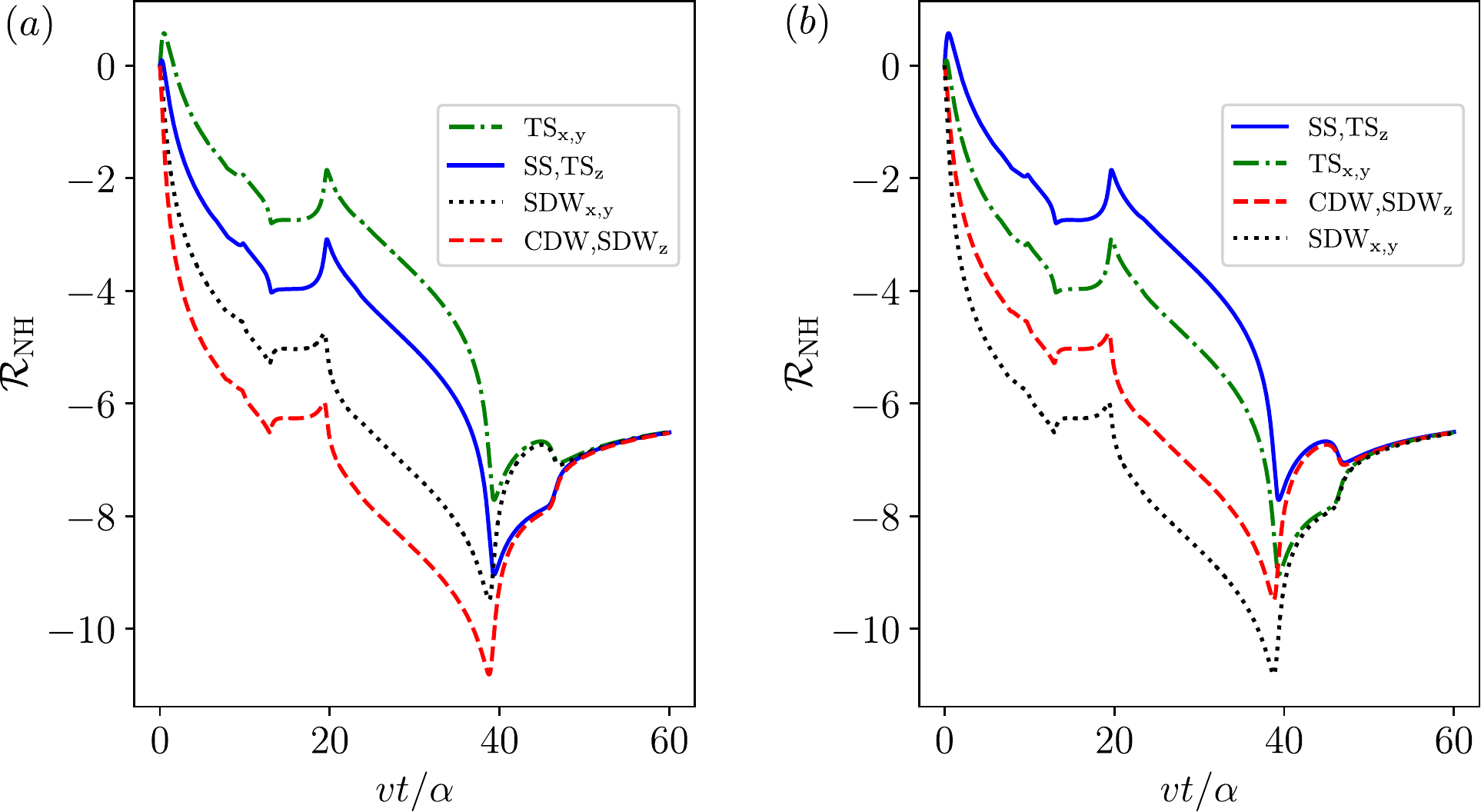}
	\caption{ We show how different types of superconducting correlations are obtained by varying anisotropy in the coupling constant. ($a$) For this plot used system parameters are: \(g_{2 \mid \mid}/v \) = 0.6 , \(g_{2 \perp}/v \) = 0.2 , \(x=100\alpha\). One can observe that the $\rm TS_{x,y}$ correlation is most dominating. ($b$) For this plot used system parameters are: \(g_{2 \mid \mid}/v \) = 0.2 , \(g_{2\perp}/v \) = 0.6  , \(x=100\alpha\). In this case, $\rm SS$ and $\rm TS_{z}$ correlations are the most dominant. Hence, modulation in anisotropy can modify the dominant behavior of the system. }\label{fig:Ani}
\end{figure*}
\begin{table}[ht!] 
 \renewcommand\thetable{I}
\renewcommand{\arraystretch}{2.}
\centering	
	\begin{tabular}{ |c||c||c||c| } 
		\hline 
		 & $\Delta_{c}$ & $\Delta_{s}$ & \text{definition of }$\mathbb{O}$  \\ 
		\hline \hline
		CDW & $-$ & $-$ & $\frac{e^{-2ik_{f}x}}{\pi\alpha}e^{i\sqrt{2}\phi_{c}(x)}\cos(\sqrt{2}\phi_{s}(x))$ \\  
		\hline 
		SS & $+$ & $-$ & $\frac{1}{\pi\alpha}e^{-i\sqrt{2}\theta_{c}(x)}\cos(\sqrt{2}\phi_{s}(x))$\\ 
		\hline
		TS$_x$ & $+$ & $+$ & $\frac{1}{\pi\alpha}e^{-i\sqrt{2}\theta_{c}(x)}\cos(\sqrt{2}\theta_{s}(x))$\\
		\hline
		TS$_y$ & $+$ & $+$ & -$\frac{1}{\pi\alpha}e^{-i\sqrt{2}\theta_{c}(x)}\sin(\sqrt{2}\theta_{s}(x))$\\ 
		\hline
		TS$_z$ & $+$ & $-$ & $\frac{e^{2ik_{f}x}}{\pi\alpha}e^{-i\sqrt{2}\theta_{c}(x)}\sin(\sqrt{2}\theta_{s}(x))$\\ 
		\hline
		SDW$_x$ & $-$ & $+$ & $\frac{e^{-2ik_{f}x}}{\pi\alpha}e^{i\sqrt{2}\phi_{c}(x)}\cos(\sqrt{2}\theta_{s}(x))$\\ 
		\hline
		SDW$_y$ & $-$ & $+$ & $-\frac{e^{-2ik_{f}x}}{\pi\alpha}e^{i\sqrt{2}\phi_{c}(x)}\sin(\sqrt{2}\theta_{s}(x))$\\ 
		\hline
		SDW$_z$ & $-$ & $-$ & $i\frac{e^{-2ik_{f}x}}{\pi\alpha}e^{i\sqrt{2}\phi_{c}(x)}\sin(\sqrt{2}\phi_{s}(x))$\\ 
		\hline  
	\end{tabular}
\caption{ We tabulate different values of $\Delta_{c,s}$ corresponding to Eqs. (\ref{eq:NH_gen},\ref{eq:H_gen}). One can obtain the expression for OCs, for any $\mathbb{O}$, using this table. We have taken the definitions of the $\mathbb{O}$ from \cite{GIAMARCHI}.  Here $k_f$ is the  Fermi momenta related to the underlying fermion model. Calculation of the identities defined in Eqs. (\ref{eq:NH_gen}, \ref{eq:H_gen} ) does not require this phase factor. Otherwise, ignoring such a phase factor ($e^{\pm 2 i k_f x}$) in the definition of $\mathbb{O}$ is not unusual \cite{FRADKIN}. } \label{Tab}
\end{table}
\section{Results and Discussion}\label{Sec:RD}
In the last section, we have shown that the scaling exponents $\mathcal{R}_{NH}$ computed for superconducting pairings dominate over that of charge and spin density waves, but in the case of a hermitian interaction quench the superconducting OCs are suppressed. We find that in the long time limit, all the exponents merge, in the case of non-Hermitian quench as shown in Fig. \ref{fig:NH}. This is in sharp contrast to the Hermitian quench counterpart, as shown in Fig. \ref{fig:Her}. We also notice spin-charge separation from Fig.~\ref{fig:NH}($b$).

One can explicitly check how SS-OC dominates over CDW-OC in the short-time limit by computing the difference between their exponents
\begin{equation}
\begin{aligned}\label{ST}
  &\mathcal{R}_{NH}^{\rm SS}-\mathcal{R}_{NH}^{\rm CDW}=
			\sum_{\nu= c,s}d_{\nu}G_{\nu}(x)\frac{g_{2\nu}}{\tilde{v}_{\nu}},
\end{aligned}
\end{equation}
where
 \begin{equation}
     \begin{aligned}\label{STd}
     \mathcal{R}_{NH}^{\eta} &= \rm log\left(\frac{\langle \mathbb{O}^{\dagger}_{\eta}(x) \mathbb{O}_{\eta}(0) \rangle_{NH}}{\langle \mathbb{O}^{\dagger}_{\eta}(x) \mathbb{O}_{\eta}(0) \rangle_{0}}\right),
     \end{aligned}
 \end{equation}
 with $\eta=\text{CDW,SS}$, and
 \begin{equation}
     \begin{aligned}\label{STdG}
     G_{\nu}(x) &= \sum_{n=0}^{\infty}\sum_{m=1}^{n+1}(-1)^{m}\mathcal{M}_{n,m} \left(\frac{g_{2\nu}^{2}}{2\tilde{v}_{\nu}^{2}}\right)^{n}  F_{\nu}(x,mt).
     \end{aligned}
 \end{equation}
We define $d_{\nu}= \Delta_{\nu}^{\rm SS}- \Delta_{\nu}^{\rm CDW}$ and note that $G_{\nu}(x)> 0 $. It is apparent from Table. \ref{Tab} that $d_{c}> 0$  and $d_{s}=0$, hence from Eq.~\eqref{ST}, the difference of exponent of the SS and CDW operator correlations is positive and thus SS dominates over CDW. Similar arguments hold for other correlations, as well. The slower decay of superconducting OC compared to the charge density wave OC remains true for the non-hermitian quench in spinless LL. In Appendix~\ref{app-D} we have shown the results for the spinless case.

If one changes the sign of the quenching (interaction) parameter to negative (i.e, an attractive interaction) then in the case of the hermitian quench this leads to an alteration in the sequence of different exponents, as shown in  Fig.~\ref{fig:Her}($b$). Interestingly, this is the same sequence observed in the case of non-Hermitian quench. This implies if we map the Bogoliubov coefficients of the non-Hermitian case to its Hermitian counterpart by using the exact mapping which has been done for the Hamiltonian, i.e. $g_{2\nu} \to -i g_{2\nu}$, then the result for Hermitian quench with negative interaction is reproduced.

Finally, we study the role of anisotropy between the quenching parameters $g_{2 \mid \mid}$ and $g_{2 \perp}$. Fig.~\ref{fig:Ani} shows that anisotropy in quenching parameters can significantly alter the exponents and may dictate which OC correlation is dominant. To understand this behavior in more detail we focus on two particular exponents $\rm TS_{x}$-OC and $\rm SS$-OC. We compute
 \begin{equation}
\begin{aligned}\label{STA}
  &\mathcal{R}_{NH}^{\rm TS_{x}}- \mathcal{R}_{NH}^{\rm SS}=	\tilde{d}_{s}G_{s}(x)\frac{g_{2s}}{\tilde{v}_{s}},
\end{aligned}
\end{equation}
where $\tilde{d}_{s}= \Delta_{s}^{\rm TS_{x}}- \Delta_{s}^{\rm SS}$ and the definitions of $\mathcal{R}_{NH}^{\eta}$ (with $\eta = \rm TS_{x}, SS$) and $G_{s}(x)$ follow from Eq. (\ref{STd}) and Eq. (\ref{STdG}) respectively. We further note, from Table. \ref{Tab}, that $\tilde{d}_{s}>0$. It is apparent from the same Table that the contribution coming from the charge sector in the difference of the exponents under study vanishes since $\Delta_{c}^{SS}=\Delta_{c}^{TS_{x}}$. These exponents merge for $g_{2 \mid \mid} = g_{2 \perp}$, which is a result of the fact that $g_{2s}=g_{2 \mid \mid} - g_{2 \perp}$ itself vanishes. If  $g_{2 \mid \mid} > g_{2 \perp}$ then, we see $g_{2s}$ is positive, so is the difference in Eq. \eqref{STA}. And hence $\rm TS_{x}$ superconducting pairing dominates over $\rm SS$. In the opposite limit i.e. $g_{2 \mid \mid} < g_{2 \perp}$, $g_{2s}$ is negative as a result of which the difference in Eq. \eqref{STA} is  negative and hence SS pairing dominate over $\rm TS_{x}$ pairing.

Other exponents can also be studied in the same manner. One would find that for $g_{2 \mid \mid} = g_{2 \perp}$, all the superconducting correlations merge but still dominate over other OCs. We note that anisotropy dictates which superconducting pairing is dominant, however, the overall dominant correlation remains superconducting, within the short time limit. For all the OCs, LL universality is retained at $t \to \infty$ \cite{Dora2020,Cazalilla2006_2}. This can be understood from the saturation attained by the exponents $\mathcal{R}_{NH,H}$. 
%

Previously, theoretical proposals have been made to realize phases of LL with dominant superconducting correlations in static hermitian systems \cite{Gefen2019,Oreg2020} under unconventional circumstances. In these referred works no attractive interaction is present and they show how superconductivity can be realized in LL platforms even if the emergence of the same is unexpected. These studies also rely upon the scaling of operator correlations to observe dominant superconductivity. In our case, one can not infer from the quenching parameter whether the interaction is attractive or repulsive since it is an imaginary number. However, we have been able to show for a non-Hermitian LL quenched out of equilibrium, superconducting behavior becomes dominant within a short time limit.

\section{Summary}\label{Sec:Sum}
We have analytically studied the problem of a non-Hermtian quantum quench in a spinful Luttinger liquid and obtained exact expressions for various operator correlations (OCs).  The exponents $\mathcal{R}_{NH}$ and $\mathcal{R}_{H}$ capture the scaling of the OCs, as shown in Eqs. (\ref{eq:NH_gen},\ref{eq:H_gen}). These exponents are crucial in understanding the system properties beyond the appearance of supersonic modes. We have explicitly discussed the results in the short-time (Eq.~(\ref{ST})) and long-time limits (Eqs.~(\ref{eq:NH_gen_LT}), \ref{eq:H_gen_LT}) by comparing the non-Hermitian quench with Hermitian quench for a spinful LL. We find that, surprisingly, for the case of non-Hermitian quench, superconducting correlations dominate over SDW or, CDW correlations (Fig.~\ref{fig:NH}), which is in stark difference to the Hermitian quench (Fig.~\ref{fig:Her}). We have also studied the effect of varying anisotropy in quenching parameters on the exponents $\mathcal{R}_{NH}$ of the OCs, as shown in Fig. \ref{fig:Ani}. With the progress made in the fields of ultracold atoms and photonic devices, which have shown potential for the implementation of non-Hermitian physics, we expect our results to be observed in similar systems where particle loss and gain can be controlled effectively~\cite{Laing2022,Rauer2016,Yosuke2020,Sponslee2019,Avila2020}. 

\section{Acknowledgment}
We acknowledge support from the SERB (Govt. of India) via sanction no. ECR/2018/001443 and CRG/2020/001803, DAE (Govt. of India ) via sanction no. 58/20/15/2019-BRNS, as well as MHRD (Govt. of India) via sanction no. SPARC/2018-2019/P538/SL.

\appendix

\begin{widetext}

\section{ Important Identities }  \label{app-A1}

 We derive various useful identities to calculate the OCs. Our main motivation, here, is to compute the expressions $ \langle e^{\pm i  \sqrt{2} \phi_{\nu}(x)}e^{\mp i \sqrt{2} \phi_{\nu}(0)}\rangle $ and $ \langle e^{\pm i  \sqrt{2} \theta_{\nu}(x)}e^{\mp i \sqrt{2} \theta_{\nu}(0)}\rangle $. These two terms are required to obtain any OC. One can observe that 
\begin{align} \label{Req_U}
	\langle e^{\pm i  \sqrt{2} \phi_{\nu}(x)}e^{\mp i \sqrt{2} \phi_{\nu}(0)}\rangle =  \langle \psi(t)\mid e^{\pm i \sqrt{2} \phi_{\nu}(x)}e^{\mp i \sqrt{2} \phi_{\nu}(0)}\mid \psi(t)\rangle = \langle \psi_{0} \mid U_{\nu}(t)  e^{\pm i \sqrt{2}( \phi_{\nu}(x,t)- \phi_{\nu}(0,t))} \mid \psi_{0}\rangle.
\end{align}
In the case of Hermitian systems, we note that the term $U_{\nu}(t)= e^{iH_{\nu}^{\dagger} t} e^{-iH_{\nu} t} $ is equal to one. This quantity solely arises due to non-Hermicity.  These terms appear in the derivation of all the OCs. We cast the above expression in the form given below
\begin{align}
	\langle \psi_{0} \mid U_{\nu}(t)  e^{\pm i \sqrt{2}( \phi_{\nu}(x,t)- \phi_{\nu}(0,t))} \mid \psi_{0}\rangle=  \langle \psi_{0} \mid U_{\nu}(t)  e^{\pm i \sqrt{2}( \phi_{\nu}^{+}(x,t)+ \phi_{\nu}^{-}(x,t))} \mid \psi_{0}\rangle.
\end{align}
We have used the relation, $\phi_{\nu}(x,t)-\phi_{\nu}(0,t)=  \phi_{\nu}^{+}(x,t)+ \phi_{\nu}^{-}(x,t)$. The Original dual fields are defined in Eq. (\ref{Dual_Fld}), in terms of $\{b_{p, \nu}^{\dagger}\}$ and $\{ b_{p, \nu} \}$ operators from which one can compute $\phi_{\nu}^{+}(x,t)$ and $\phi_{\nu}^{-}(x,t)$ fields as
\begin{align}
	\phi_{\nu}^{+}(x,t)= -i\frac{\pi}{L}\sum_{p>0}\left(\frac{L p}{2\pi}\right)^{\frac{1}{2}}\frac{1}{p}e^{-\frac{\alpha p}{2}}(u_{p, \nu}^{*}(t)+ v_{p, \nu}(t))\left[(e^{-i p x}- 1)b_{p, \nu}^{\dagger}- (e^{i p x}- 1)b_{-p, \nu}^{\dagger}\right],\\
	\phi_{\nu}^{-}(x,t)= -i\frac{\pi}{L}\sum_{p>0}\left(\frac{L p}{2\pi}\right)^{\frac{1}{2}}\frac{1}{p}e^{-\frac{\alpha p}{2}}(v_{p, \nu}(t))- u_{p, \nu}(t))\left[(e^{i p x}- 1)b_{p, \nu}- (e^{-i p x}- 1)b_{-p, \nu}\right].
\end{align}
We can also write
\begin{align}\label{cal_1}
	\langle \psi_{0} \mid U_{\nu}(t)  e^{\pm i \sqrt{2}( \phi_{\nu}^{+}(x,t)+ \phi_{\nu}^{-}(x,t))} \mid \psi_{0}\rangle=  \langle \psi_{0} \mid U_{\nu}(t)  e^{\pm i \sqrt{2}( \phi_{\nu}^{+}(x,t)}e^{i\pm\sqrt{2} \phi_{\nu}^{-}(x,t))}e^{\left[\phi_{\nu}^{+}(x,t), \phi_{\nu}^{-}(x,t)\right]} \mid \psi_{0}\rangle. 
\end{align}
The commutator in the above expression is given by
\begin{align}
	\left[\phi_{\nu}^{+}(x,t), \phi_{\nu}^{-}(x,t)\right]= \sum_{p>0}\frac{4\pi}{L p}e^{-\alpha p}\sin^{2}\left(\frac{p x}{2}\right)(u^{*}_{p, \nu}(t)v_{p, \nu}(t)- v_{p, \nu}(t)u_{p, \nu}(t)+ 2\mid v_{p, \nu}(t)\mid^{2}- 1).
\end{align}
We note that \(e^{\pm i\sqrt{2}\phi_{\nu}^{-}(x,t)}\mid \psi_{0}\rangle= \mid \psi_{0}\rangle\) and as a result we get
\begin{align}
	\langle \psi_{0}\mid U_{\nu}(t) = \langle \psi_{0}\mid e^{i H_{\nu}^{\dagger} t}e^{-i H_{\nu} t}= \langle \psi_{0}\mid \prod_{p>0} e^{\mathcal{C}_{-\nu}(p,t)\mathcal{K}_{p, \nu}^{-}+ \frac{\mathcal{C}_{o \nu}(p,t)}{2}}= N_{\nu}(t) \langle \psi_{0}\mid \prod_{p>0} e^{\mathcal{C}_{-\nu}(p,t)\mathcal{K}_{p, \nu}^{-}},
\end{align}
where \(\langle \psi(t)\mid \psi(t)\rangle= \mathcal{N} (t)  \) as defined in Eq. (\ref{OCNH}). We have decoupled the contribution coming from spin and charge sectors by writing \(\mathcal{N} (t) = \prod_{\nu=c,s} \mathcal{N}_{\nu}(t)  \). So calculating Eq. (\ref{cal_1}) reduces to computation of $\langle \psi_{0} \mid U_{\nu}(t)  e^{\pm i \sqrt{2}( \phi_{\nu}^{+}(x,t)} \mid \psi_{0}\rangle e^{\left[\phi_{\nu}^{+}(x,t), \phi_{\nu}^{-}(x,t)\right]}$. After doing some algebraic manipulations one can write it in the form below
\begin{align}\nonumber
	&\langle \psi_{0} \mid U_{\nu}(t)  e^{\pm i \sqrt{2}( \phi_{\nu}^{+}(x,t)} \mid \psi_{0}\rangle e^{\left[\phi_{\nu}^{+}(x,t), \phi_{\nu}^{-}(x,t)\right]} \\ \nonumber
	=&  \langle \psi_{0}\mid \prod_{p>0} \sum_{mn}\frac{(\mathcal{C}_{-\nu}(p,t)k_{p, \nu}^{-})^{m}}{m!} \frac{\left(\pm \sqrt{\frac{\pi}{L p}}e^{-\frac{\alpha p}{2}}(u_{p, \nu}^{*}(t)+ v_{p, \nu}(t))\left((e^{-i p x}- 1)b_{p, \nu}^{\dagger}- (e^{i p x}- 1)b_{-p, \nu}^{\dagger}\right)\right)^{n}}{n!} \mid \psi_{0}\rangle
	\\&~ e^{\left[\phi_{\nu}^{+}(x,t), \phi_{\nu}^{-}(x,t)\right]} ~, \\\nonumber
	=& \prod_{p>0} \sum_{m}(-\mathcal{C}_{-\nu}(p,t))^{m}\left( \frac{\pi}{L p}e^{-\alpha p}(u_{p, \nu}^{*}(t)+ v_{p, \nu}(t))^{2}\right)^{m} \left(4\sin^{2}\left(\frac{px}{2}\right)\right)^{m}  e^{\left[\phi_{\nu}^{+}(x,t), \phi_{\nu}^{-}(x,t)\right]} ~.
\end{align}
We can thus write the operator identities as
\begin{align}\label{Exp_ph_1}
\langle e^{\pm i  \sqrt{2} \phi_{\nu}(x)}e^{\mp i \sqrt{2} \phi_{\nu}(0)}\rangle 
&=\mathcal{N}_{\nu}(t)e^{-\sum_{p>0}\frac{4\pi}{L p}e^{-\alpha p}\sin^{2}\left(\frac{p x}{2}\right)((u_{p, \nu}^{*}(t)+ v_{p, \nu}(t))^{2}\mathcal{C}_{-\nu}(p,t)- (u_{p, \nu}^{*}(t)v_{p, \nu}(t)- v_{p, \nu}(t)u_{p, \nu}(t)+ 2\mid v_{p, \nu}(t) \mid^{2}- 1))}.
\end{align}
Now we put the value of  \(\mathcal{C}_{-\nu} \) in the above equation from Eq. (\ref{Cm}) and use the expression of \(u_{p, \nu}(t)\) and \(v_{p, \nu}(t)\) from Eq. (\ref{coeff}) to reduce the above expression in the form 
\begin{align} \label{Exp_ph_2}
\langle e^{\pm i  \sqrt{2} \phi_{\nu}(x)}e^{\mp i \sqrt{2} \phi_{\nu}(0)}\rangle 
=\mathcal{N}_{\nu}(t) e^{-\sum_{p>0}\frac{4\pi}{L p}e^{-\alpha p}\sin^{2}\left(\frac{p x}{2}\right)\tilde{v}^2_{\nu}\left(\frac{1+ \frac{2g_{2\nu}}{\tilde{v}_{\nu}}\sin(\tilde{v}_{\nu}pt)\cos(\tilde{v}_{\nu}pt)}{\tilde{v}_{\nu}^{2}-2g_{2\nu}^{2}\sin^{2}(\tilde{v}_{\nu}pt)}\right)}.
\end{align}
Now, consider the argument of exponential in the above equation
\begin{align}
\nonumber
\sum_{p>0}\frac{4\pi}{L p}e^{-\alpha p}\sin^{2}\left(\frac{p x}{2}\right)\tilde{v}_{\nu}^{2}\left(\frac{1+ \frac{2g_{2\nu}}{\tilde{v}_{\nu}}\sin(\tilde{v}_{\nu}pt)\cos(\tilde{v}_{\nu}pt)}{\tilde{v}_{\nu}^{2}- 2g_{2\nu}^{2}\sin^{2}(\tilde{v}_{\nu}pt)}\right)&= \sum_{p>0}\frac{4\pi}{L p}e^{-\alpha p}\sin^{2}\left(\frac{p x}{2}\right)\left(\frac{\tilde{v}_{\nu}^{2}}{\tilde{v}_{\nu}^{2}- 2g_{2\nu}^{2}\sin^{2}(\tilde{v}_{\nu}pt)}\right)\\
&+ \sum_{p>0}\frac{4\pi}{L p}e^{-\alpha p}\sin^{2}\left(\frac{p x}{2}\right)\left(\frac{\tilde{v}_{\nu}^{2}\frac{2g_{2\nu}}{\tilde{v}_{\nu}}\sin(\tilde{v}_{\nu}pt)\cos(\tilde{v}_{\nu}pt)}{\tilde{v}_{\nu}^{2}- 2g_{2\nu}^{2}\sin^{2}(\tilde{v}_{\nu}pt)}\right).
\end{align}
 We proceed further by converting the sum over $p$ to an integral over $p$ such that in the thermodynamic limit $L \to \infty$ one gets
\begin{align}
\nonumber
&\sum_{p>0}\frac{4\pi}{L p}e^{-\alpha p}\sin^{2}\left(\frac{p x}{2}\right)\left(\frac{\tilde{v}_{\nu}^{2}}{\tilde{v}_{\nu}^{2}- 2g_{2\nu}^{2}\sin^{2}(\tilde{v}_{\nu}pt)}\right)= \int_{0}^{\infty}\frac{dp}{p}e^{-\alpha p}(1- \cos(px))\sum_{n=0}\left(\frac{2g_{2\nu}^{2}}{\tilde{v}_{\nu}^{2}}\sin^{2}(\tilde{v}_{\nu}pt)\right)^{n}\\
&=\sum_{n=0}\left(\frac{\frac{2g_{2\nu}^{2}}{\tilde{v}_{\nu}^{2}}}{4(-1)}\right)^{n}\sum_{r=0}^{2n}(-1)^{2n-r} \frac{2n!}{r!(2n-r)!}\frac{1}{2}\log\left[1+ \frac{x^{2}}{(\alpha+ 2it\tilde{v}_{\nu}(n-r))^{2}}\right].
\end{align}
And finally, we write
\begin{align}
\nonumber \label{Exp_ph_3}
\sum_{p>0}\frac{4\pi}{L p}e^{-\alpha p}\sin^{2}\left(\frac{p x}{2}\right)\tilde{v}_{\nu}^{2}&\left(\frac{1+ \frac{2g_{2\nu}}{\tilde{v}_{\nu}}\sin(\tilde{v}_{\nu}pt)\cos(\tilde{v}_{\nu}pt)}{\tilde{v}_{\nu}^{2}- 2g_{2\nu}^{2}\sin^{2}(\tilde{v}_{\nu}pt)}\right) \nonumber \\ &= \frac{\tilde{v}_{\nu}}{v_{-\nu}}D_{\nu}(x,0)+ 2\sum_{n=1}^{\infty}\frac{\tilde{v}_{\nu}}{v_{-\nu}}\left(\frac{-g_{2\nu}^{2}}{v^{2}+ \tilde{v}_{\nu}v_{-\nu}}\right)^{n}D_{\nu}(x,nt) \nonumber\\
&+ \frac{g_{2\nu}}{\tilde{v}_{\nu}}\sum_{n=0}^{\infty}\left(\frac{g_{2\nu}^{2}}{2\tilde{v}_{\nu}^{2}}\right)^{n}\sum_{m=1}^{m=n+1}(-1)^{m}\left[{}^{2n}\mathbf{C}_{n-m-1}- {}^{2n}\mathbf{C}_{n-m+1}\right]F_{\nu}(x,mt),
\end{align}
where the forms of \(\tilde{v}_{\nu}\) , \(v_{-\nu}\), \(D_{\nu}(x,t)\) and  \(F_{\nu}(x,t)\) are defined in the main text. One can check that the terms of the form $\langle e^{\pm i  \sqrt{2} \phi_{\nu}(x)}e^{\pm i \sqrt{2} \phi_{\nu}(0)}\rangle $ can be written as
\begin{align} \label{vanp}
	\nonumber
	\langle e^{\pm i \sqrt{2} \phi_{\nu}(x)}e^{\pm i \sqrt{2} \phi_{\nu}(0)}\rangle 
	&=\mathcal{N}_{\nu}(t) e^{-\sum_{p>0}\frac{4\pi}{L p}e^{-\alpha p}\cos^{2}\left(\frac{p x}{2}\right)(u_{p, \nu}^{*}(t)+ v_{p, \nu}(t))^{2}\mathcal{C}_{-\nu}(p,t)} \nonumber \\& ~~~~~~~~\times e^{\sum_{p>0}\frac{4\pi}{L p}e^{-\alpha p}\cos^{2}\left(\frac{p x}{2}\right)(u_{p, \nu}^{*}(t)v_{p, \nu}(t)- v_{p, \nu}(t)u_{p, \nu}(t)+ 2\mid v_{p, \nu}(t) \mid^{2}- 1)}. 
\end{align}
It is analytically and numerically checked that these terms vanish and have no contribution in the correlation functions. We proceed with $\theta_{\nu}$ fields in the same manner to obtain
\begin{align}
\nonumber \label{Exp_th_1}
\langle e^{\pm i  \sqrt{2} \theta_{\nu}(x)}e^{\mp i \sqrt{2} \theta_{\nu}(0)}\rangle 
& =\mathcal{N}_{\nu}(t)e^{\sum_{p>0}\frac{4\pi}{L p}e^{-\alpha p}\sin^{2}\left(\frac{p x}{2}\right)(u_{p, \nu}^{*}(t)- v_{p, \nu}(t))^{2}\mathcal{C}_{-\nu}(p,t)} \nonumber \\ &~~~~~~~~\times e^{-\sum_{p>0}\frac{4\pi}{L p}e^{-\alpha p}\sin^{2}\left(\frac{p x}{2}\right)(u_{p, \nu}^{*}(t)v_{p, \nu}(t)- v_{p, \nu}(t)u_{p, \nu}(t)- 2\mid v_{p, \nu}(t) \mid^{2}+ 1)}.
\end{align}
Which turns out to be
\begin{align}
\nonumber \label{Exp_th_2}
\langle e^{\pm i  \sqrt{2} \theta_{\nu}(x)}e^{\mp i \sqrt{2} \theta_{\nu}(0)}\rangle 
= &\mathcal{N}_{\nu}(t)  \exp\left(-\frac{\tilde{v}_{\nu}}{v_{-\nu}}D_{\nu}(x,0)- 2\sum_{n=1}^{\infty}\frac{\tilde{v}_{\nu}}{v_{-\nu}}\left(\frac{-g_{2\nu}^{2}}{v^{2} + \tilde{v}_{\nu}v_{-\nu}}\right)^{n}D_{\nu}(x,n.t) \right) \\& ~~~~\times
\exp\left( \frac{g_{2\nu}}{\tilde{v}_{\nu}}\sum_{n=0}^{\infty}\left(\frac{g_{2\nu}^{2}}{2\tilde{v}_{\nu}^{2}}\right)^{n}\sum_{m=1}^{m=n+1}(-1)^{m}\left[{}^{2n}\mathbf{C}_{n-m-1}- {}^{2n}\mathbf{C}_{n-m+1}\right]F_{\nu}(x,m.t) \right). 
\end{align}
We also calculate
\begin{align}
\nonumber 
\langle e^{\pm i  \sqrt{2} \theta_{\nu}(x)}e^{\pm i \sqrt{2} \theta_{\nu}(0)}\rangle 
&=\mathcal{N}_{\nu}(t)e^{\sum_{p>0}\frac{4\pi}{L p}e^{-\alpha p}\cos^{2}\left(\frac{p x}{2}\right)(u_{p, \nu}^{*}(t)- v_{p, \nu}(t))^{2}\mathcal{C}_{-\nu}(p,t)} \nonumber\\&~~~~~~~~\times e^{-\sum_{p>0}\frac{4\pi}{L p}e^{-\alpha p}\cos^{2}\left(\frac{p x}{2}\right)(u_{p, \nu}^{*}(t)v_{p, \nu}(t)- v_{p, \nu}(t)u_{p, \nu}(t)- 2\mid v_{p, \nu}(t) \mid^{2}+ 1)}. \label{vant}
\end{align}
It has been mathematically checked that Eq. (\ref{vant}) vanishes similarly to Eq. (\ref{vanp}). So far we have computed the quantities which are crucial in the calculation of different OCs. We keep in mind that OCs for CDW and SS are shown, in the next section, in detail and all the other OCs can be calculated in a similar manner. 
\section{ non-Hermitian quench } \label{app-A2}
The study of various OCs relies upon detailed computation of different expressions furnished in Table. \ref{Tab}, over time evolved ground state $| \psi (t) \rangle  = e^{-i H t} | \psi_0 \rangle $. The boson operators $\{ b_{p,\nu} \}$ and $\{ b_{p,\nu}^{\dagger} \}$ are related to the dual fields as per Eq. (\ref{Dual_Fld}). The time evolution of these boson operators takes place according to Eq. (\ref{PHTE}). One specific OC is a particular combination of the dual boson fields of Eq. (\ref{Dual_Fld}). In order to find out the OCs, we need to find out different expectations (taken over $| \psi (t) \rangle $ ) of exponentiated $\phi_{\nu}$ and $\theta_{\nu}$ fields. We focus on two particular cases, where the CDW-OC requires the calculation of the product of exponentials defined with $\phi_{\nu}$ fields and SS-OC where the same is done with  $\theta_{\nu}$ fields. These two OCs capture all the technical details required to compute other OCs.
\subsection{ OC for CDW } \label{app-A3}
The CDW correlator takes the form (Table. \ref{Tab})
\begin{align}
	\langle \mathbb{O}^{\dagger}_{\rm CDW}(x) \mathbb{O}_{\rm CDW}(0) \rangle_{NH}
	&= \frac{e^{2ik_{f}x}}{\pi^{2}\alpha^{2}}\frac{\langle \psi(t) \mid e^{-i\sqrt{2}\phi_{c}(x)}\cos(\sqrt{2}\phi_{s}(x))e^{i\sqrt{2}\phi_{c}(0)}\cos(\sqrt{2}\phi_{s}(0))\mid \psi(t)\rangle}{\langle \psi(t)\mid \psi(t)\rangle} ~.
\end{align}
We can write the OC for CDW in a compact form as
\begin{align}
	\langle \mathbb{O}^{\dagger}_{\rm CDW}(x) \mathbb{O}_{\rm CDW}(0) \rangle_{NH}
	&= \frac{e^{2ik_{f}x}}{2 \pi^{2}\alpha^{2}}\frac{\langle e^{-i \sqrt{2}\phi_{c}(x)}e^{i\sqrt{2}\phi_{c}(0)}\rangle (\langle e^{i \sqrt{2}\phi_{s}(x)}e^{i\sqrt{2}\phi_{s}(0)}\rangle+ \langle e^{-i \sqrt{2}\phi_{s}(x)}e^{i\sqrt{2}\phi_{s}(0)}\rangle)}{\mathcal{N}_{c}(t)\mathcal{N}_{s}(t)}.
\end{align}
In writing the above equation we have used the fact that \( \langle e^{i \sqrt{2}\phi_{s}(x)}e^{i\sqrt{2}\phi_{s}(0)}\rangle= \langle e^{-i \sqrt{2}\phi_{s}(x)}e^{-i\sqrt{2}\phi_{s}(0)}\rangle\) and \(\langle e^{-i \sqrt{2}\phi_{s}(x)}e^{i\sqrt{2}\phi_{s}(0)}\rangle= \langle e^{i \sqrt{2}\phi_{s}(x)}e^{-i\sqrt{2}\phi_{s}(0)}\rangle\). So using Eq. (\ref{Exp_ph_1}) the non-zero part of the above expression can be written as
\begin{align}
	\nonumber
	\langle \mathbb{O}^{\dagger}_{\rm CDW}(x) \mathbb{O}_{\rm CDW}(0) \rangle_{NH}  = 
	& \frac{e^{2ik_{f}x}}{2\pi^{2}\alpha^{2}}e^{-\sum_{p>0} \frac{4\pi}{L p}e^{-\alpha p}\sin^{2}\left(\frac{p x}{2}\right)(u_{p,c}^{*}(t)+ v_{p,c}(t))^{2}\mathcal{C}_{-c}(p,t)} \\&~~~~~~~~~
	\times e^{\sum_{p>0}\frac{4\pi}{L p}e^{-\alpha p}\sin^{2}\left(\frac{p x}{2}\right)(u_{p,c}^{*}(t)v_{p,c}(t)- u_{p,c}(t)v_{p,c}(t)+ 2\mid v_{p,c}(t)\mid^{2}-1)}\mathcal{L}_{s}(p,t),
\end{align}
where \(\mathcal{L}_{s}(p,t)\) is given by
\begin{align}
	\mathcal{L}_{s}(p,t) &= e^{-\sum_{p>0} \frac{4\pi}{L p}e^{-\alpha p}\sin^{2}\left(\frac{p x}{2}\right)(u_{p,s}^{*}(t)+ v_{p,s}(t))^{2}\mathcal{C}_{-s}(p,t)}e^{\sum_{p>0}\frac{4\pi}{L p}e^{-\alpha p}\sin^{2}\left(\frac{p x}{2}\right)(u_{p,s}^{*}(t)v_{p,s}(t)- u_{p,s}(t)v_{p,s}(t)+ 2\mid v_{p,s}(t)\mid^{2}-1)}.
\end{align}
All the terms can be combined to obtain
\begin{align}
	\nonumber
	\langle \mathbb{O}^{\dagger}_{\rm CDW}(x) \mathbb{O}_{\rm CDW}(0) \rangle_{NH}  = 
	&\frac{e^{2ik_{f}x}}{2\pi^{2}\alpha^{2}}e^{-\sum_{p>0} \frac{4\pi}{L p}e^{-\alpha p}\sin^{2}\left(\frac{p x}{2}\right)(u_{p,c}^{*}(t)+ v_{p,c}(t))^{2}\mathcal{C}_{-c}(p,t)}e^{-\sum_{p>0} \frac{4\pi}{L p}e^{-\alpha p}\sin^{2}\left(\frac{p x}{2}\right)(u_{p,s}^{*}(t)+ v_{p,s}(t))^{2}\mathcal{C}_{-s}(p,t)}\\
	&~~~~~\times e^{\sum_{p>0}\frac{4\pi}{L p}e^{-\alpha p}\sin^{2}\left(\frac{p x}{2}\right)(u_{p,s}^{*}(t)v_{p,s}(t)- u_{p,s}(t)v_{p,s}(t)+ 2\mid v_{p,s}(t)\mid^{2}-1)} \nonumber \\&~~~~~\times e^{\sum_{p>0}\frac{4\pi}{L p} e^{-\alpha p}\sin^{2}\left(\frac{p x}{2}\right)(u_{p,c}^{*}(t)v_{p,c}(t)- u_{p,c}(t)v_{p,c}(t)+ 2\mid v_{p,c}(t)\mid^{2}-1)}.
\end{align}
We next use the identities defined in Eqs. (\ref{Exp_ph_2}-\ref{Exp_ph_3}) to write
\begin{equation}
	\begin{aligned} 
		&\langle \mathbb{O}^{\dagger}_{\rm CDW}(x) \mathbb{O}_{\rm CDW}(0) \rangle_{NH} \\&= \frac{e^{2ik_{f}x}}{2\pi^{2}\alpha^{2}}   \exp\left(-\frac{\tilde{v}_{c}}{v_{-c}}D_{c}(x,0)- 2\sum_{n=1}^{\infty}\frac{\tilde{v}_{c}}{v_{-c}}\left(\frac{-g_{2c}^{2}}{v^{2}+ \tilde{v}_{c}v_{-c}}\right)^{n}D_{c}(x,nt)- \frac{g_{2c}}{\tilde{v}_{c}}\sum_{n=0}^{\infty}\sum_{m=1}^{n+1}(-1)^{m} \mathcal{M}_{n,m} F_{c}(x,mt) \left(\frac{g_{2c}^{2}}{2\tilde{v}_{c}^{2}}\right)^{n} \right)  \\
		&~~~~~~~~~\times\exp\left(-\frac{\tilde{v}_{s}}{v_{-s}}D_{s}(x,0)-2\sum_{n=1}^{\infty}\frac{\tilde{v}_{s}}{v_{-s}}\left(\frac{-g_{2s}^{2}}{v^{2}+ \tilde{v}_{s}v_{-s}}\right)^{n}D_{s}(x,nt)- \frac{g_{2s}}{\tilde{v}_{s}}\sum_{n=0}^{\infty} \sum_{m=1}^{n+1}(-1)^{m} \mathcal{M}_{n,m} F_{s}(x,mt) \left(\frac{g_{2s}^{2}}{2\tilde{v}_{s}^{2}}\right)^{n} \right) ,
	\end{aligned}	
\end{equation}
where $\mathcal{M}_{n,m}$ is defined in the main text. If we divide the correlation by its bare part and convert the summations over $p$ to integrals over $p$ as discussed in the previous section we would get
\begin{align}
	\nonumber
	\frac{\langle \mathbb{O}^{\dagger}_{\rm CDW}(x) \mathbb{O}_{\rm CDW}(0) \rangle_{NH}}{\langle \mathbb{O}^{\dagger}_{\rm CDW}(x) \mathbb{O}_{\rm CDW}(0) \rangle_{0}}
	=& \exp\left(\left(1-\frac{\tilde{v}_{c}}{v_{-c}}\right)D_{c}(x,0)- 2\sum_{n=1}^{\infty}\frac{\tilde{v}_{c}}{v_{-c}}\left(\frac{-g_{2c}^{2}}{v^{2}+ \tilde{v}_{c}v_{-c}}\right)^{n}D_{c}(x,n.t)\right) \\\nonumber
	\times &\exp\left(-\frac{g_{2c}}{\tilde{v}_{c}}\sum_{n=0}^{\infty}\left(\frac{g_{2c}^{2}}{2\tilde{v}_{c}^{2}}\right)^{n}\sum_{m=1}^{m=n+1}(-1)^{m} \mathcal{M}_{n,m} F_{c}(x,m.t)\right) \\\nonumber
	\times  & \exp\left(\left(1-\frac{\tilde{v}_{s}}{v_{-s}}\right)D_{s}(x,0)- 2\sum_{n=1}^{\infty}\frac{\tilde{v}_{s}}{v_{-s}}\left(\frac{-g_{2s}^{2}}{v^{2}+ \tilde{v}_{s}v_{-s}}\right)^{n}D_{s}(x,n.t)\right) \\
	\times  & \exp\left(- \frac{g_{2s}}{\tilde{v}_{s}}\sum_{n=0}^{\infty}\left(\frac{g_{2s}^{2}}{2\tilde{v}_{s}^{2}}\right)^{n}\sum_{m=1}^{m=n+1}(-1)^{m} \mathcal{M}_{n,m} F_{s}(x,m.t)\right).
\end{align}
\subsection{ OC for SS }  \label{app-A4}
This part of  the Appendix is dedicated to the computation of the SS-OC. The analytical calculation of the SS-OC is similar to that of the CDW-OC but with $\theta_{\nu}$ fields. Using Table. \ref{Tab}, one can write the SS-OC as
\begin{align}
	\langle \mathbb{O}^{\dagger}_{\rm SS}(x) \mathbb{O}_{\rm SS}(0) \rangle_{NH}
	= \frac{1}{\pi^{2}\alpha^{2}}\frac{\langle \psi(t) \mid e^{i\sqrt{2}\theta_{c}(x)}\cos(\sqrt{2}\phi_{s}(x))e^{-i\sqrt{2}\theta_{c}(0)}\cos(\sqrt{2}\phi_{s}(0))\mid \psi(t)\rangle}{\langle \psi(t)\mid \psi(t)\rangle}.
\end{align}
Using equation Eq. (\ref{Exp_th_1}) we can write
\begin{align}
	\langle \mathbb{O}^{\dagger}_{\rm SS}(x) \mathbb{O}_{\rm SS}(0) \rangle_{NH}  & = \frac{1}{2\pi^{2}\alpha^{2}}e^{\sum_{p>0} \frac{4\pi}{L p}e^{-\alpha p}\sin^{2}\left(\frac{p x}{2}\right)(u_{p,c}^{*}(t)- v_{p,c}(t))^{2}\mathcal{C}_{-c}(p,t)} \nonumber \\ &~~~~~~~~~\times e^{-\sum_{p>0}\frac{4\pi}{L p}e^{-\alpha p}\sin^{2}\left(\frac{p x}{2}\right)(u_{p,c}^{*}(t)v_{p,c}(t)- u_{p,c}(t)v_{p,c}(t)- 2\mid v_{p,c}(t)\mid^{2}+1)}\mathcal{L}_{s}(p,t),
\end{align}
where \(\mathcal{L}_{s}(p,t)\) is given by
\begin{align}
	\mathcal{L}_{s}(p,t) &= e^{-\sum_{p>0} \frac{4\pi}{L p}e^{-\alpha p}\sin^{2}\left(\frac{p x}{2}\right)(u_{p,s}^{*}(t)+ v_{p,s}(t))^{2}\mathcal{C}_{-s}(p,t)}e^{\sum_{p>0}\frac{4\pi}{L p}e^{-\alpha p}\sin^{2}\left(\frac{p x}{2}\right)(u_{p,s}^{*}(t)v_{p,s}(t)- u_{p,s}(t)v_{p,s}(t)+ 2\mid v_{p,s}(t)\mid^{2}-1)}.
\end{align}
Next, one can follow the similar steps used to compute CDW-OC, elaborated in the Appendix (\ref{app-A3}), but with fields $\theta_{\nu}$ to obtain
\begin{equation}
	\begin{aligned} 
		&\langle \mathbb{O}^{\dagger}_{\rm SS}(x) \mathbb{O}_{\rm SS}(0) \rangle_{NH} \\ &= \frac{1}{2\pi^{2}\alpha^{2}}  \exp\left(-\frac{\tilde{v}_{c}}{v_{-c}}D_{c}(x,0)- 2\sum_{n=1}^{\infty}\frac{\tilde{v}_{c}}{v_{-c}}\left(\frac{-g_{2c}^{2}}{v^{2}+ \tilde{v}_{c}v_{-c}}\right)^{n}D_{c}(x,nt)+ \frac{g_{2c}}{\tilde{v}_{c}}\sum_{n=0}^{\infty}\sum_{m=1}^{n+1}(-1)^{m} \mathcal{M}_{n,m} F_{c}(x,mt) \left(\frac{g_{2c}^{2}}{2\tilde{v}_{c}^{2}}\right)^{n} \right) \\&~~~~~~~~~\times \exp\left(-\frac{\tilde{v}_{s}}{v_{-s}}D_{s}(x,0)-2\sum_{n=1}^{\infty}\frac{\tilde{v}_{s}}{v_{-s}}\left(\frac{-g_{2s}^{2}}{v^{2}+ \tilde{v}_{s}v_{-s}}\right)^{n}D_{s}(x,nt)- \frac{g_{2s}}{\tilde{v}_{s}}\sum_{n=0}^{\infty}\sum_{m=1}^{n+1}(-1)^{m}\mathcal{M}_{n,m} F_{s}(x,mt) \left(\frac{g_{2s}^{2}}{2\tilde{v}_{s}^{2}}\right)^{n} \right).
	\end{aligned}	
\end{equation}
One has to use Eq. (\ref{Exp_th_2}) to calculate the above expression. If we divide the correlation with its bare part and convert $p$-sum to $p$-integral, then we get
\begin{align}
	\nonumber
	\frac{\langle \mathbb{O}^{\dagger}_{\rm SS}(x) \mathbb{O}_{\rm SS}(0) \rangle_{NH}}{\langle \mathbb{O}^{\dagger}_{\rm SS}(x) \mathbb{O}_{\rm SS}(0) \rangle_{0}}
	=& \exp\left(\left(1-\frac{\tilde{v}_{c}}{v_{-c}}\right)D_{c}(x,0)- 2\sum_{n=1}^{\infty}\frac{\tilde{v}_{c}}{v_{-c}}\left(\frac{-g_{2c}^{2}}{v^{2}+ \tilde{v}_{c}v_{-c}}\right)^{n}D_{c}(x,n.t)\right)\\ \nonumber
	\times &\exp\left(\frac{g_{2c}}{\tilde{v}_{c}}\sum_{n=0}^{\infty}\left(\frac{g_{2c}^{2}}{2\tilde{v}_{c}^{2}}\right)^{n}\sum_{m=1}^{m=n+1}(-1)^{m} \mathcal{M}_{n,m} F_{c}(x,m.t)\right) \\\nonumber
	\times  & \exp\left(\left(1-\frac{\tilde{v}_{s}}{v_{-s}}\right)D_{s}(x,0)- 2\sum_{n=1}^{\infty}\frac{\tilde{v}_{s}}{v_{-s}}\left(\frac{-g_{2s}^{2}}{v^{2}+ \tilde{v}_{s}v_{-s}}\right)^{n}D_{s}(x,n.t)\right)\\
	\times  & \exp\left(-\frac{g_{2s}}{\tilde{v}_{s}}\sum_{n=0}^{\infty}\left(\frac{g_{2s}^{2}}{2\tilde{v}_{s}^{2}}\right)^{n}\sum_{m=1}^{m=n+1}(-1)^{m} \mathcal{M}_{n,m} F_{s}(x,m.t)\right).
\end{align}

\section{ Hermitian quench} \label{app-A5}
In this part of the Appendix, we briefly discuss the calculation of different OCs, in the case of Hermitian interaction quench. It has been noted that the calculation does not require pseudo-Heisenberg representation and everything can be calculated in the Heisenberg picture. We begin by writing down the two most important entities for the calculation
\begin{align}
	\langle e^{\pm i  \sqrt{2} \phi_{\nu}(x)}e^{\mp i \sqrt{2} \phi_{\nu}(0)}\rangle &=  \langle \psi (t) \mid e^{\pm i \sqrt{2} \phi_{\nu}(x)}e^{\mp i \sqrt{2} \phi_{\nu}(0)}\mid \psi (t) 
	= \langle \psi_{0} \mid  e^{i\pm \sqrt{2}( \phi_{\nu}(x,t)- \phi_{\nu}(0,t))} \mid \psi_{0}\rangle, \\
	\langle e^{\pm  i \sqrt{2} \theta_{\nu}(x)}e^{\mp  i \sqrt{2} \theta_{\nu}(0)}\rangle &=  \langle \psi (t)\mid e^{\pm i \sqrt{2} \theta_{\nu}(x)}e^{\mp i \sqrt{2} \theta_{\nu}(0)}\mid \psi (t)\rangle 
	= \langle \psi_{0} \mid  e^{i\pm \sqrt{2}( \theta_{\nu}(x,t)- \theta_{\nu}(0,t))} \mid \psi_{0}\rangle.
\end{align}
Following the similar arguments and steps as that of the non-Hermitian case we can write
\begin{align}
	\langle e^{\pm i  \sqrt{2} \phi_{\nu}(x)}e^{\mp i\sqrt{2} \phi_{\nu}(0)}\rangle  = \exp\left(-\sum_{p>0}\frac{4\pi}{L p}e^{-\alpha p}\sin^{2}\left(\frac{p x}{2}\right)(u^{*}_{p, \nu}(t)v_{p, \nu}(t)+ v_{p, \nu}^{*}(t)u_{p, \nu}(t)+ 2\mid v_{p, \nu}(t)\mid^{2}+ 1)\right).
\end{align}
The terms of the form $\langle e^{\pm i  \sqrt{2} \phi_{\nu}(x)}e^{\pm i\sqrt{2} \phi_{\nu}(0)}\rangle $ vanish in a manner similar to the  non-Hermitian case. We can further  write
\begin{align}
	\langle e^{\pm i  \sqrt{2} \phi_{\nu}(x)}e^{\mp i\sqrt{2} \phi_{\nu}(0)}\rangle = \exp\left(-\left(1+ \frac{g_{2\nu}(g_{2\nu}-v)}{v_{-\nu}^{2}}\right)D_{\nu}(x,0)+ \frac{g_{2\nu}(g_{2\nu}-v)}{v_{-\nu}^{2}}D_{\nu}(x,t)\right).
\end{align}
For $\theta_{\nu}$ fields we can write
\begin{align}
	\langle e^{\pm i \sqrt{2} \theta_{\nu}(x)}e^{\mp i\sqrt{2} \theta_{\nu}(0)}\rangle &= \exp\left( \sum_{p>0}\frac{4\pi}{L p}e^{-\alpha p}\sin^{2}\left(\frac{p x}{2}\right)(u^{*}_{p, \nu}(t)v_{p, \nu}(t)+ v_{p, \nu}^{*}(t)u_{p, \nu}(t)- 2\mid v_{p, \nu}(t)\mid^{2}- 1)\right) \nonumber\\
	&= \exp\left(-\left(1+ \frac{g_{2\nu}(g_{2\nu}+v)}{v_{-\nu}^{2}}\right)D_{\nu}(x,0)+ \frac{g_{2\nu}(g_{2\nu}+v)}{v_{-\nu}^{2}}D_{\nu}(x,t)\right).
\end{align}
Next, we write down the explicit expression for the Bogoliubov coefficients used in our calculation, for the Hermitian case
\begin{equation}
\begin{aligned} \label{coeff_H}
	u_{p, \nu}(t)&= \cos(\tilde{v}_{\nu}| p | t)- \frac{i v}{\tilde{v}_{\nu}}\sin(\tilde{v}_{\nu}| p | t),\\
	v_{p, \nu}(t)&= \frac{i g_{2\nu}}{\tilde{v}_{\nu}}\sin(\tilde{v}_{\nu} 
	| p | t). 
\end{aligned}	
\end{equation}
We have used the above identities to calculate the OCs for the Hermitian case. In this regard, one is referred to Fig. \ref{fig:Her}($a$), to observe the effect of repulsive interaction quench in OCs. In Fig. \ref{fig:Her}($b$) we have shown the results for \( g_{2\nu} \to - g_{2 \nu} \).
\subsubsection*{ \rm OC \it for \rm CDW \it and \rm SS }
In case of Hermitian quench, $\langle \psi(t)\mid \psi(t)\rangle=1$. Following the same techniques used for the calculations done in the case of non-Hermitian quench, we compute the exact expressions for CDW and SS correlators. The CDW-OC takes the form
\begin{align}
	\nonumber
	\langle \mathbb{O}^{\dagger}_{\rm CDW}(x) \mathbb{O}_{\rm CDW}(0) \rangle_{H}
	&= \frac{e^{2ik_{f}x}}{\pi^{2}\alpha^{2}}\frac{\langle \psi(t) \mid e^{-i\sqrt{2}\phi_{c}(x)}\cos(\sqrt{2}\phi_{s}(x))e^{i\sqrt{2}\phi_{c}(0)}\cos(\sqrt{2}\phi_{s}(0))\mid \psi(t)\rangle}{\langle \psi(t)\mid \psi(t)\rangle}\\
	&= \frac{e^{2ik_{f}x}}{2 \pi^{2}\alpha^{2}} \langle e^{-i \sqrt{2}\phi_{c}(x)}e^{i\sqrt{2}\phi_{c}(0)}\rangle (\langle e^{i \sqrt{2}\phi_{s}(x)}e^{i\sqrt{2}\phi_{s}(0)}\rangle+ \langle e^{-i \sqrt{2}\phi_{s}(x)}e^{i\sqrt{2}\phi_{s}(0)}\rangle).
\end{align}
We extract the non-zero part from the above expression as
\begin{align}
	\nonumber
	\langle \mathbb{O}^{\dagger}_{\rm CDW}(x) \mathbb{O}_{\rm CDW}(0) \rangle_{H}  = 
	&\frac{e^{2ik_{f}x}}{2\pi^{2}\alpha^{2}}e^{-\sum_{p>0}\frac{4\pi}{L p}e^{-\alpha p}\sin^{2}\left(\frac{p x}{2}\right)(u_{p,c}^{*}(t)v_{p,c}(t)+ u_{p,c}(t)v_{p,c}^{*}(t)+ 2\mid v_{p,c}(t)\mid^{2}+1)}\\
	&~~~~~~\times
	e^{-\sum_{p>0}\frac{4\pi}{L p}e^{-\alpha p}\sin^{2}\left(\frac{p x}{2}\right)(u_{p,s}^{*}(t)v_{p,s}(t)+ u_{p,s}(t)v_{p,s}^{*}(t)+ 2\mid v_{p,s}(t)\mid^{2}+1)}.
\end{align}
If we divide the correlation with its bare part and convert $p$-sum to $p$-integral, then we get
\begin{align}
	\frac{\langle \mathbb{O}^{\dagger}_{\rm CDW}(x) \mathbb{O}_{\rm CDW}(0) \rangle_{H}}{\langle \mathbb{O}^{\dagger}_{\rm CDW}(x) \mathbb{O}_{\rm CDW}(0) \rangle_{0}}&=  e^{-\left(\frac{g_{2c}(g_{2c}-v)}{v_{-c}^{2}}\right)(D_{c}(x,0)-D_{c}(x,t)) } e^{-\left(\frac{g_{2s}(g_{2s}-v)}{v_{-s}^{2}}\right)(D_{s}(x,0)- D_{s}(x,t)) }.
\end{align}
We can calculate the above quantity for SS-OC, In the same manner, as
\begin{align}
	\frac{\langle \mathbb{O}^{\dagger}_{\rm SS}(x) \mathbb{O}_{\rm SS}(0) \rangle_{H}}{\langle \mathbb{O}^{\dagger}_{\rm SS}(x) \mathbb{O}_{\rm SS}(0) \rangle_{0}} &=  e^{-\left(\frac{g_{2c}(g_{2c}+v)}{v_{-c}^{2}}\right)(D_{c}(x,0)- D_{c}(x,t) )} e^{-\left(\frac{g_{2s}(g_{2s}-v)}{v_{-s}^{2}}\right)(D_{s}(x,0)-D_{s}(x,t)) }.
\end{align}

\begin{figure*}
	\centering
	\includegraphics[width=.8\textwidth]{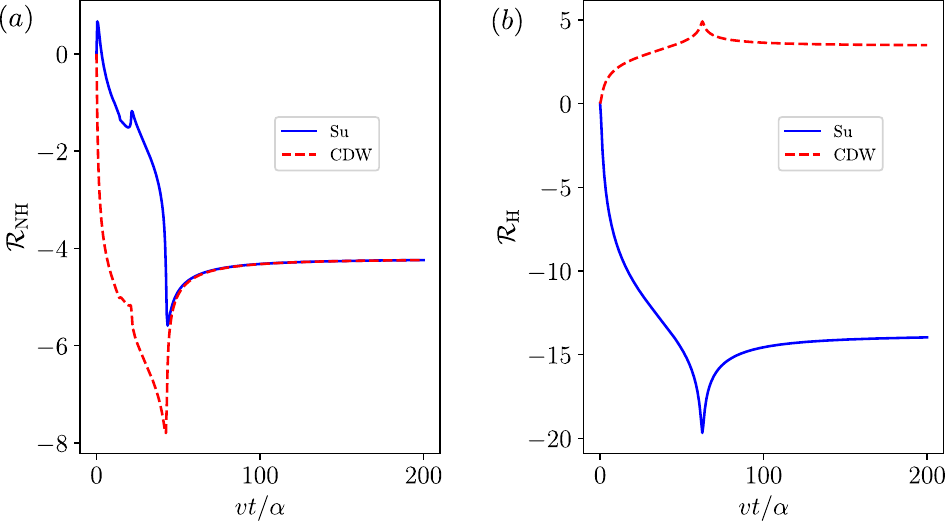}
	\caption{ We show results for spinless LL ($a$) We show OCs for non-Hermitian case. For this plot used system parameters are: \(g_{2}/v \) = 0.6,  \(x=100\alpha\). One can observe that the $\rm Su$ correlation is the most dominating. ($b$) We show OCs for Hermitian case. For this plot used system parameters are: \(g_{2 }/v \) = 0.6 ,  \(x=100\alpha\). In this case, $\rm CDW$ correlation is the most dominant. We observe that supersonic mode is present in case of non-Hermitian quench as shown in panel ($a$) but absent in panel ($b$). }\label{fig:SL}
\end{figure*}
\section{ Results for spinless LL }  \label{app-D}
One can compute OCs, under non-Hermitian and Hermitian sudden quench, for the spinless case as well (see Fig.~\ref{fig:SL}). We have obtained the exponents $\mathcal{R}_{\rm NH}$ and $\mathcal{R}_{\rm H}$ for the spinless LL, which we show below. The calculation follows in the same manner as we have elaborated in Appendix  (\ref{app-A1}-\ref{app-A5}).  We note that for this case, Su and CDW stand for superconductor and charge density wave respectively. We define the operators corresponding to Su and CDW OCs as \cite{FRADKIN,GIAMARCHI}
\begin{equation}
    \begin{aligned}
        \mathbb{O}_{\rm Su}(x)&= \frac{e^{-2i\theta(x)}}{\pi\alpha} ~,~
         \mathbb{O}_{\rm CDW}(x)&= \frac{e^{-2i\phi(x)}}{\pi\alpha}.
    \end{aligned}
\end{equation}
Here the dual bosonic fields $\phi(x)$ and $\theta(x)$ are defined in the same way as we have done in Eq. (\ref{Dual_Fld}) but without the $\nu$ index.

\end{widetext}

\end{document}